# J-PET analysis framework for the prototype TOF-PET detector

Wojciech Krzemień[1], Michał Silarski[1], Karol Stoła[1], Damian Trybek[1], Tomasz Bednarski[1], Piotr Białas[1], Eryk Czerwiński[1], Łukasz Kapłon[1,2], Andrzej Kochanowski[2], Grzegorz Korcyl[1], Jakub Kowal[1], Paweł Kowalski[3], Tomasz Kozik[1], Marcin Molenda[2], Paweł Moskal[1], Szymon Niedźwiecki[1], Marek Pałka[1], Monika Pawlik[1], Lech Raczyński[3], Zbigniew Rudy[1], Piotr Salabura[1], Neha Gupta Sharma[1], Artur Słomski[1], Jerzy Smyrski[1], Adam Strzelecki[1], Wojciech Wiślicki[3], Marcin Zieliński[1] and Natalia Zoń[1]

[1] Institute of Physics, Jagiellonian University, Kraków, Poland

[2] Faculty of Chemistry, Jagiellonian University, Kraków, Poland

[3] Świerk Computing Centre, National Centre for Nuclear Research, 05-400 Otwock-Świerk, Poland

Corresponding author: Michal Silarski, Jagiellonian University - Department of Nuclear Physics

Reymonta 4 Kraków 30-059, Poland

E-mail: michal.silarski@uj.edu.pl

**Keywords**: analysis framework, STRIP-PET

**Abstract:** Novel time-of-flight positron emission tomography (TOF-PET) scanner solutions demand, apart from the state of the art detectors, software for fast processing of the gathered data, monitoring of the whole scanner and reconstruction of the PET image. In this article we present an analysis framework for the novel STRIP-PET scanner developed by the J-PET collaboration in the Institute of Physics of the Jagiellonian University. This software is based on the ROOT package used in many particle physics experiments.

# Introduction

Positron emission tomography (PET) imagining appears to be one of the most dynamically developing field of medical sciences. Apart from the state of the art detectors for annihilation gamma quanta registration, PET imagining demands software for fast processing of the gathered data, monitoring of the whole scanner and reconstruction of the PET image. Typical PET scanners are built out of hundreds of detection modules which measure both, the energy and the arrival time of gamma quanta. This results in a huge amount of data to be stored and processed. For novel TOF-PET (time-of-flight PET) devices [1-5] the number of channels to be read out could be much higher due to e.g., reading detection modules from both sides, as in the case of STRIP-PET scanner [6] being developed by the J-PET collaboration in the Institute of Physics of the Jagiellonian University.

The processing of the data collected with the PET scanner is complex and it must be performed in several steps. The process starts with the raw TDC (time-to-digital converter) and ADC (amplitude-to-digital converter) data been provided by the front-end electronics, which are then combined into signals and translated to hit positions in the individual scintillator bars. Finally, the signals are matched to form the line of response (LOR), that is the line between the detected gamma quanta coming from the annihilation of the positron emitted from the body of the patient. The set of LORs is then used as an input for the image reconstruction procedure. Moreover, at many stages of the data processing the adequate calibration procedure must be applied. Additional issues arise from the fact that many different algorithms and approaches are used and tested in parallel at the development phase of the prototype.

The novel TOF-PET scanner developed by the J-PET collaboration will consists of hundreds of detection modules based on long plastic scintillator strips read out from both sides by photomultipliers. Registered signals are probed in the voltage domain at several levels by a newly developed dedicated electronics working in a trigger-less mode [7-8], which results in a large data flow. The probing will allow for reconstruction of the gamma quanta interaction point based on the signal shapes. For this, special reconstruction procedures will be applied, and many possible methods can be conceived ( see e.g. [9]).

In order to address the above mentioned needs we have developed a flexible analysis framework which serves as a backbone system for the reconstruction algorithms and calibration procedures used during the data processing and standardize the common operations, e.g: input/output process, access to the detector geometry parameters and more.

The general idea of the framework was adapted from the existing solutions used in the field of high-energy and nuclear physics like ALICERoot used by the ALICE collaboration [10] or RootSorter used by ANKE and WASA collaborations [11]. However, the J-PET detection system is relatively simple compared to the full detection system used in experimental physics, like WASA, typically composed of many sub-detectors of different types, therefore we could restrict our design to a much more lightweight architecture.

## Architecture and data flow

The J-PET analysis framework is written in C++ using the object-oriented approach. It is based on the ROOT libraries [12]. The quality of the code is assured by the automatized set of unit tests. We use the BOOST unit test framework [13]. The documentation of the code is generated by Doxygen [14]. The main components of the framework are presented in Figure 1.

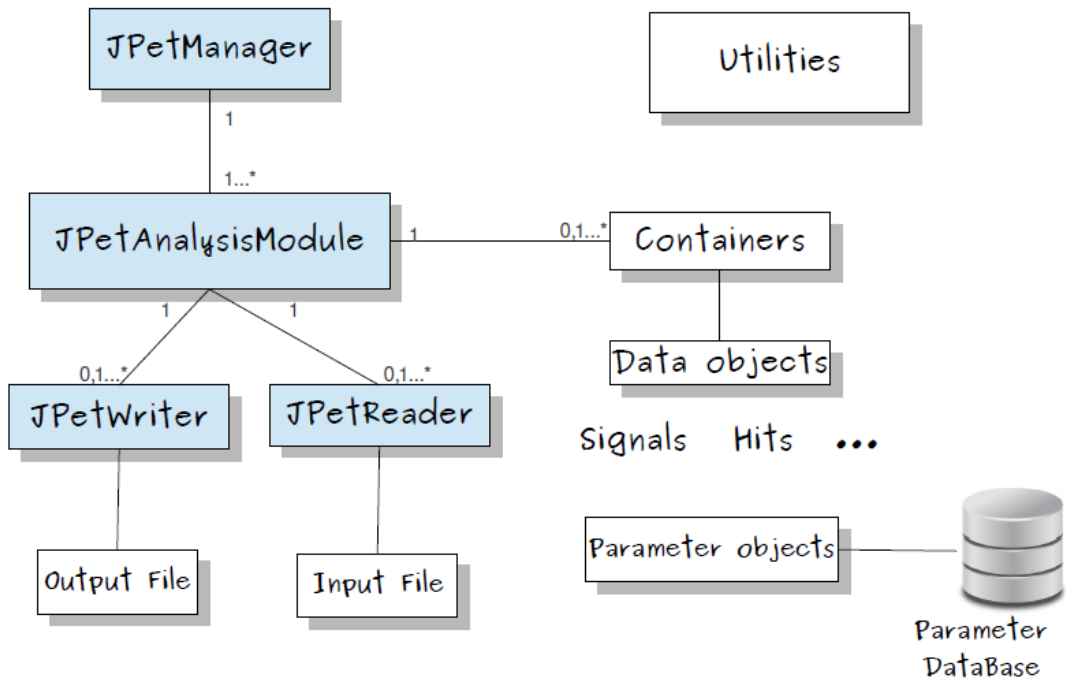

Figure 1 General architecture of the J-PET framework.

The JPetAnalysisModule could be for example the module for monitoring and calibration of the STRIP-PET scanner.

The JPetManager is the main part of the system. Its task is to steer the data flow and to synchronize the work of the components. It can also register JPetAnalysisModules which will perform particular computation tasks like calibration procedures, hit matching, LOR reconstructions, etc. Each JPetAnalysisModule component is a separate module that can be activated or deactivated depending on the current needs. This solution provides an easy way to incorporate new modules in the existing code.

The J-Pet framework provides predefined interfaces for the input/output operation at every level of the data processing via the JPetWriter and JPetReader components. The modules provide the common set of methods that handle operations of reading from a file and writing to a file. It also permits to save the processed objects at intermediate levels of the data processing (see Figure 2).

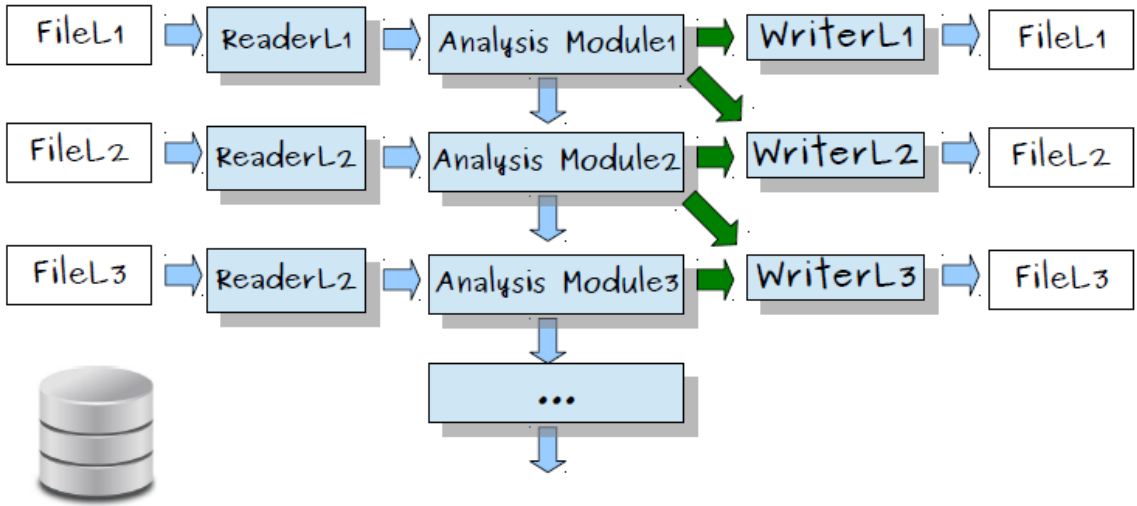

Fig. 2 Schematic data flow in the reconstruction process. The data processing is performed by the Analysis Modules, which could be, for example, calibration or hit matching modules, and which send the data to the next level of the reconstruction. At every level, the intermediate results can be saved using the JPetWriter modules. Moreover, additional data can be read, if necessary, via the JPetReader interfaces. The final result of this procedure is the set of LORs which is then used as an input to the image reconstruction algorithms.

An important part of the J-PET system is the parameter database, which contains information about the geometry configurations, the DAQ setups, specific run setups and HV settings that can be important during the data reconstruction and analysis. This information is implemented as a PostgreSQL database. The J-PET framework will also automatize the access to the parameter database by providing the corresponding parameter data interface. The development is in progress.

## Conclusions

We have developed the analysis framework for the prototype STRIP-PET detection system. This software is based on the open source package ROOT used in many experiments in particle physics. Since we are dealing with a large data flow, our software was optimized in view of computation time and computational resources utilization. The framework is being used for the off-line analysis and tests but a part of it will be also used as an on-line reconstruction module of the larger software system,

PetController, which is developed for steering of the whole PET measurements.

## Acknowledgments

We acknowledge technical and administrative support by M. Adamczyk, T. Gucwa-Ryś, A. Heczko, M. Kajetanowicz, G. Konopka-Cupiał, J. Majewski, W. Migdał, A. Misiak, and the financial support by the Polish National Center for Development and Research through grant INNOTECH-K1/IN1/64/159174/NCBR/12, the Foundation for Polish Science through MPD programme and the EU and MSHE Grant No. POIG.02.03.00-161 00-013/09.

## References

1. Shehad N. N., Athanasiades A., Martin Ch. S., Sun L., Lacy J. L. Novel Lead-Walled Straw PET Detector for Specialized Imaging Applications. Nuclear Science Symposium Conference Record, 2005 IEEE 2005;4: 2895- 2898.

2. Lacy J. L. , Martin Ch. S, Armendarez L. P. High sensitivity, low cost PET using lead-walled straw detectors. Nucl. Instrum. Meth. A 2001;471:88–93.

3. Blanco A., Couceiro M., Crespo P., Ferreira N. C., Ferreira Marques R., Fonte P., Lopes L., Neves J. A. Efficiency of RPC detectors for whole-body human TOF-PET. Nucl. Instrum. Meth. A 2009;602: 780–783

4. Moskal P., Salabura P., Silarski M., Smyrski J., Zdebik J., Zieliński M. Novel detector systems for the Positron Emission Tomography. Bio-Algorithms and Med-Systems 2012;7:73-78; e-print arXiv:1305.5187.

5. Moskal P., Bednarski T, Białas P., Ciszewska M., Czerwiński E., Heczko A., et al. TOF-PET detector concept based on organic scintillators. Nucl. Med. Rev. 2012;15:C81-C84; e-print arXiv:1305.5559.

6. Moskal P., Bednarski T, Białas P., Ciszewska M., Czerwiński E., Heczko A., et al. STRIP-PET: a novel detector concept for the TOF-PET scanner. Nucl. Med. Rev. 2012;15:C68-C69; e-print arXiv:1305.5562.

7. Pałka M., P., Bednarski T., Bałas P., Czerwiński E., Kapłon Ł., Kochanowski A., et al. A novel method based solely on FPGA units enabling measurement of time and charge of analog signals in Positron Emission Tomography. Bio-Algorithms and Med-Systems 2013, accompanying article in this issue.

8. Korcyl G., Moskal P., Bednarski T., Bałas P., Czerwiński E., Kapłon Ł., et al. Trigger-less and reconfigurable data acquisition system for positron emission tomography. Bio-Algorithms and Med-Systems 2013, accompanying article in this issue.

9. Raczyński L., Kowalski P., Bednarski T., Bałas P., Czerwiński E., Kapłon Ł., et. al. Application of Compressive Sensing Theory for the Reconstruction of Signals in Plastic Scintillators. Acta Phys. Polon. B Proceed. Suppl. 2013,6:1121-1127 e-print arXiv:1310.1612.

10. AliRoot documentation. Available at: http://aliweb.cern.ch/Offline/AliRoot/Manual.html.

11. Hejny V., Hartmann M., Mussgiller A. RootSorter: A New Analysis Framework for ANKE. IKP Annual Report 2002.

12. Brun R., Rademakers F. ROOT - An Object Oriented Data Analysis Framework. Nucl. Instrum. Meth. A 1997,389:81-86.

13. BOOST. Available at: http://www.boost.org/.

14. Doxygen. Available at: http://www.stack.nl/~dimitri/doxygen/.